# Full-Head Segmentation of MRI with Abnormal Brain Anatomy: Model and Data Release

Andrew M Birnbaum[a], Adam Buchwald[b], Peter Turkeltaub[c], Adam Jacks[d], George Carr[a], Shreya Kannan[a], Yu Huang[a,e], Abhisheck Datta[e], Lucas C Parra[a], Lukas A Hirsch[a]*

[a]The City College of New York, Department of Biomedical Engineering, New York, NY, USA
[b]New York University, New York, NY, USA
[c]Georgetown University, Department of Neurology, Washington, DC, USA
[d]University of North Carolina School of Medicine, Department of Allied Health Sciences, Chapel Hill, NC, USA
[e]Soterix Medical, New York, NY, USA

**Abstract**

**Purpose:** The goal of this work was to develop a deep network for whole-head segmentation including clinical MRIs with abnormal anatomy, and compile the first public benchmark dataset for this purpose. We collected 98 MRIs with volumetric segmentation labels for a diverse set of human subjects including normal, as well as abnormal anatomy in clinical cases of stroke and disorders of consciousness.

**Approach:** Training labels were generated by manually correcting initial automated segmentations for skin/scalp, skull, CSF, gray matter, white matter, air cavity and extracephalic air. We developed a "MultiAxial" network consisting of three 2D U-Net that operate independently in sagittal, axial and coronal planes and are then combined to produce a single 3D segmentation.

**Results:** The MultiAxial network achieved a test-set Dice scores of 0.88±0.04 (median ± interquartile range) on whole head segmentation including gray and white matter. This compared to 0.86 ± 0.04 for Multipriors and 0.79 ± 0.10 for SPM12, two standard tools currently available for this task. The MultiAxial network gains in robustness by avoiding the need for coregistration with an atlas. It performed well in regions with abnormal anatomy and on images that have been de-identified. It enables more accurate and robust current flow modeling when incorporated into ROAST, a widely-used modeling toolbox for transcranial electric stimulation.

**Conclusions:** We are releasing a new state-of-the-art tool for whole-head MRI segmentation in abnormal anatomy, along with the largest volume of labeled clinical head MRIs including labels for non-brain structures. Together the model and data may serve as a benchmark for future efforts.

*Index Terms*—Magnetic Resonance Imaging, Convolutional Neural Network, Segmentation, Stroke

## 1 Introduction

Magnetic Resonance Imaging (MRI) of the brain is used extensively in clinical practice, in medical research and in neuroscience. Volumetric assessment of brain anatomy plays an important role in these areas. In a clinic, measuring the extent of a tumor is important for planning of surgery and radiation therapy[1] as well as monitoring of tumor progression and treatment response[2]. Measuring brain volume is important also in diagnosis of conditions such as Alzheimer's disease[3], traumatic brain injury[4], hydrocephalus[5] and stroke[6]. Measuring brain volume implies delineating cerebro-spinal fluid (CSF), which fills up the space left behind by a deteriorating brain. In chronic stroke in particular, the lesioned brain tissue is cleared up and the space is filled in with CSF in as little as 6 months from the injury[7]. In neuroscience, measuring the volume of gray and white matter in different brain areas also plays an important role in explaining brain function[8]. Finally, the idiosyncratic differences in anatomy across individuals can have large effects on encephalographic brain signals[9] and transcranial brain stimulation[10]. Therefore current flow modeling, as used in ROAST[11,12], Brainstorm[13], and SimNIBS[14], all required accurate automatic segmentation of the human head. Yet, existing tools are largely limited to brain segmentation, and fail for clinical cases with abnormal anatomy. For all these use cases, it would be important to have a tool that can reliably segment, not just the brain, but the entire head in the presence of lesions and abnormal anatomy.

*Address all correspondence to Lukas A Hirsch, lukashirsch@gmail.com

Currently, automatic segmenters such as SPM[15], FSL[16], and DeepMedic[17], all perform well on individuals with normal anatomy. There are also tools, which perform detailed segmentation of different areas within the brain, such as hippocampus, midbrain, cerebellum[18] or even detailed perceptions of cortical tissue[19]. Unfortunately these tools struggle with abnormal anatomy or lesions. A number of tools have been developed to identify tumors[20] or lesions related to MS[21] and ALS[22]. There are also some existing methods to delineate chronic stroke lesions such as LINDA[23]. However, to our knowledge, these methods do not carefully segment non-brain tissues, even in normal anatomies, beyond "skull stripping". In the context of current flow modeling, skull, CSF and non-brain soft tissue are equally important given their drastic difference in electric conductivity[24] (skull: 0.02 S/m, CSF: 1.71 S/m, scalp: 0.41 S/m). Current flow models are important to compute "lead fields" that are used in current source reconstructions in MEG/EEG, or to target specific brain areas with TES/TMS. Segmentation tools developed in these contexts, such as in ROAST[11,12], BrainStorm[13], and SimNIBS[14], work well on normal anatomy, but again, fail in the presence of lesions. Therefore, there is a need for an automatic MRI brain segmentation tool that will perform reliably in cases with abnormal anatomy.

Traditional methods, such as SPM and FSL rely on relatively simple intensity features and a tissue probability map (TPM). We improved on this approach by extracting more complex features using a convolutional network and adding morphological priors. When trained on lesioned anatomies, this "MultiPriors" network[25] improved performance compared to SPM and FSL. Unfortunately, aligning the TPM to the MRI has remained a persistent challenge in the presence of abnormal anatomy.

Here we developed a deep network for segmentation that does not require alignment with an atlas. Instead, the network was trained on a larger corpus of clinical scans with chronic stroke or disorder of consciousness (typically resulting from traumatic brain injury). The network is based on a standard 2D U-Net[26] that efficiently and reliably segments MRIs into 7 tissue classes. The 2D U-Net is applied in axial, sagittal and coronal planes following a well-established approach for efficient 3D processing in AI [27–30]. In this work we focus on the task of segmenting the whole head in subjects with abnormal brain anatomy, which remains a challenging problem. An important part of this new work was to generate the corresponding manually annotated dataset for training and testing. We compared performance to our MultiPriors model and SPM for whole head segmentation. This shows state-of-the-art performance for automatic segmentation for abnormal anatomies. Focusing on brain tissue only, we also compared to newer deep networks, such as Brainchop[31] and SynthSeg[32] and found that SynthSeg perform better on normal anatomy, but worse abnormal anatomies. Foundation models for segmentation, such as SAM[33] and MED-SAM[34] required extensive manual input and were not further explored. The data used here corresponds to one of the largest labeled data for full head MRI in terms of labeled pixels. We release the network and data that was used for testing in the hope that it can serve as a benchmark for future efforts.

## 2 METHODS

**2.1** *Head MRI from patients with chronic abnormalities and healthy subjects*

The data was collected from four different institutions from patients with different clinical conditions (See Table 1). All scans are T1-weighted MRI from a total of 98 individuals as follows: 4 scans from healthy subjects (Gender: 4M/ 0F, Ethnicity: 1 Asian, 3 White (not Latino), Age: 30-50 ) obtained on a 3T Siemens Trio scanner (Erlangen, Germany); 50 patients with chronic aphasia stroke collected at Georgetown University (Gender: 18M/ 10F, Ethnicity: 1 Asian, 8 Black,  17 White (not Latino), 1 White/Latino, 1 Unknown, Age: 41-75) imaged on a Siemens Trio 3T scanner and University of North Carolina, Chapel Hill (Gender: 16M/ 5F/ 1 Unknown, Ethnicity:  8 Black, 14 White (not Latino), Age: 44-75), imaged on a 3T Siemens Biograph mMR scanner; 34 subjects with disorders of consciousness collected at the Pitie-Salpetriere University Hospital in Paris, imaged on a 3T General Electric Signa system Milwaukee, WI (Demographics not disclosed) and 7 patients with chronic aphasia and apraxia of speech subsequent to stroke collected at New York University (Gender: 8 M/ 2W, Ethnicity: 1 Black, 1 Asian, 7 White (not Latino), 1 White/Latino Age: 38-85), imaged on a 3T Siemens Prisma scanner. Patients had the lesion or injuries occur at least 6 months prior to the MRI scan, at which point the lesioned brain area is largely replaced by CSF. All scans have isotropic resolution of 1mm. Data was provided by the clinical sites in de-identified form and the IRB of the City University of New York approved data analysis of pre-existing data under an exempt review.

**2.2** *Creation of segmentation labels*

For all available MRIs we generated volumetric segmentations labeling all voxels into one of seven possible categories: Background (extracephalic air), air cavities, white matter (WM), gray matter (GM), cerebrospinal fluid (CSF), bone and non-brain soft tissue. These labels were obtained for the full 3D volume using semi-manual segmentation. Specifically, healthy (N=3+1) and aphasia stroke heads (N=50) were first segmented automatically using SPM (version 8), touched up by an in-house Matlab script[35], and then manually corrected using scanIP (Synopsys, Mountain View, CA). Manual correction focused on the stroke lesions and boundaries between CSF, gray matter and skull boundary. One head from a previous study[36] had been fully segmented manually using ScanIP.

For the stroke patients with apraxia (N=10) and patients with Disorders of Consciousness (N=34), the head MRIs were segmented using Multipriors[25] through ROAST[12]  which utilized SPM12 for coregistration with a TPM. Post-processing steps in Python addressed common segmentation errors, such as mislabeled background within the skull and sinus, bone touching brain matter instead of CSF, skin mislabeled as CSF, and air mislabeled as bone. A custom Python script corrected these errors by relabeling bone in contact with the brain as CSF, converting background within skin regions to bone, and reclassifying sinus cavities outside the head as background. Despite these automated corrections, manual relabeling was essential to address residual errors, ensuring accurate and anatomically reliable segmentations. Manual adjustments using MRIcron's[37] drawing tools corrected significant anatomical mistakes, particularly in regions where skin, CSF, and air were mislabeled.

Additional, manual corrections were applied to the data before training and release. Cranial bone and CSF continuities were completed, and grey matter, white matter, and CSF boundaries were improved using filters to reduce noise and unevenness. Unrealistic or fragmented outer skin mapping was corrected with filters or manual unpainting. Incomplete spinal columns were extended manually and surrounded with CSF. Non-sinus air was relabeled as background, and mislabeled contents within eyes were corrected to CSF.

Portions of the data used in this study have been publicly released with de-identified and defaced or original MRIs, along with manually-corrected labels version 2 (Table 1, and see Code and Data availability).

**Table 1: Overview of the labels and MRIs used for training, testing and public release.**

| Subjects | Labels | Train | Test | Data release (on Kaggle) |
|---|---|---|---|---|
| Normal | 1 manual +3 SPM + manual correction | 4 | 3 (cross validated) | Original MRI + training labels |
| Chronic Aphasia Stroke | SPM + manual correction | 50 | 50 (cross validated) | Defaced MRI + training labels |
| Chronic Apraxia Stroke | MultiPriors + manual correction | 10 | 0 | Defaced MRI + training labels |
| Disorders of Consciousness | MultiPriors + manual correction | 34 | 0 | None |

**2.3** *Anonymization for public data release*

For anonymization purposes, the reface tool of AFNI[38,39] was used, which provides three methods of face removal or replacement: deface, reface, and reface+. The reface option adds an average face to the MRI, while reface+ also generates a uniform head shape. However, reface+ generates unrealistic scalp/skull anatomy and will not be considered here.

**2.4** *Data preprocessing*

Harmonization of the input involved resampling the MRI to 1 mm isotropic resolution if needed, reoriented to 'RAS' (right - anterior - superior), and resizing to 256x256x256 voxels through padding/cropping. Intensity normalization was done by dividing by each image's 95 percentile.

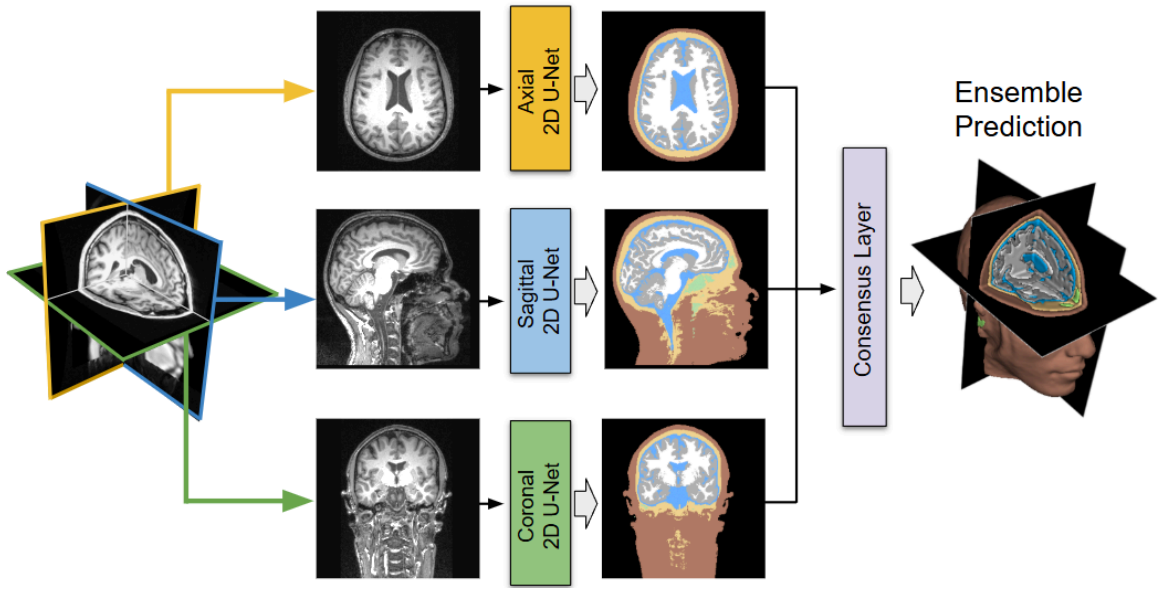

**Figure 1: Overview of the MultiAxial model for segmentation of head MRI.** An input T1 weighted MRI is processed in each axis (axial, sagittal and coronal) by a separate 2D U-Net, each of which output probabilities per pixel for belonging to one of seven classes: Background (black), Air/Sinus cavities (light green), White Matter (white), Gray Matter (gray), Cerebrospinal fluid (CSF, light blue), bone (light yellow) and skin (brown). The probabilities from each model are merged into a single volume by the consensus layer, consisting of a single 3D convolutional layer.

**2.5** *Model architecture and training*

We utilized three 2D U-Nets to process 2D slices of the MRI in each of the three orthogonal axes: Sagittal, Coronal and Axial. The three models have identical architecture but their own set of weights. Each model outputs the probability of belonging to one of 7 possible tissue classes for each pixel in the volume. We use a single 3D convolutional layer of kernel size 1 to merge these three probabilities into a single output, referred to as a "consensus layer" (Fig. 1, and Fig. S1).

The input dimension is fixed to 256x256 for all models, processed by 6 convolutional blocks each consisting of 2 convolutional layers with kernel size (3,3) and ReLU activation[40], followed by a downsampling layer with a pool size of (2,2). The upsampling is done by 6 convolutional blocks with a transposed convolutional layer instead. Each of the three models has a total of approximately 1 million trainable parameters. The network also takes as input the coordinates of the image to provide for spatial awareness as a matrix with three channels, encoding each of the x,y,z axes. This spatial information is concatenated before the last two layers of the network. During training, the network uses the Adam optimizer[41] with a learning rate of 1e-5 and a dice loss function for multiclass classification.[42,43].

**2.6** *Model selection and training*

A total of 98 heads were used for training and testing, with 53 used for testing overlapping between train and test using cross-validation (see Table 1). Each cross-validation run split 98 heads into training+validation+testing (two runs split as 76+12+10 and three runs as 75+12+11). Slices that contain only background from the boundaries of each MRI were removed from the training data. The three models for each direction were trained separately. Once trained, the consensus layer was initialized to perform a majority vote and further trained from there (Fig. S1). Selection of model and training hyperparameters was done using the validation set for performance evaluation.

**2.7** *Evaluation and benchmarks*

We compare the performance of the Multiaxial model on the test set (Table 1) with the following public tools and pre-trained models: SPM, Multipriors, Brainchop and SynthSeg. For the evaluation metric we use the Dice score, which measures overlap between masks ranging from 0 (no overlap) to 1 (perfect overlap). We average Dice scores across tissues and report the median and mean Dice score across subjects.

We used the Multipriors model as integrated in the software 'ROAST'. This convolutional neural network takes as an input a TPM, which needs to be aligned with the input MRI. Within this software (in version 3.0), the TPM alignment is performed by SPM12. We also tested alignment with SPM8, which proved to be more robust for abnormal anatomies with default parameters.

Brainchop segmentations on our data with the pre-trained model were obtained following instructions provided on the Brainchop GitHub[44] repository. Brainchop was installed with Pip and ran with the terminal command: "brainchop -i <input> -o <output> -m -subcortical." Following recommendations by Brainchop developers, we used the robust subcortical-plus-gray-white-matter model, which outputs a segmentation with 14/18 classes within the brain, corresponding to parcellation of different brain regions. To obtain a segmentation of only white and gray matter instead, we mapped these classes to two classes by overlapping the Brainchop output with our ground-truth labels, determining which regions best correspond to either white or gray matter. Small isolated connected components were removed using the connected-components-3d library.

SynthSeg segmentations on our data with the pre-trained model were obtained following instructions provided on the SynthSeg Github[45] repository. We cloned the repository, created a virtual environment with the required dependencies, downloaded the model files, and ran the following terminal command: "python ./scripts/commands/SynthSeg_predict.py --i <input> --o <output> --robust." We used the robust model, which outputs a segmentation with 33 classes within the brain. The same mapping process as used for Brainchop was applied to classify the output into gray matter, white matter, CSF or background. For both Brainchop and SynthSeg Dice scores were calculated only for gray matter and white matter.

**2.8** *Out-of-sample evaluation*

We used an external dataset, Mindboggle[46], for an out-of-sample comparison to SynthSeg. For completeness, we also report results for Brainchop, although this is the training data for that mode[47]. The Mindboggle dataset consists of T1 MRI images (N=101 subjects compiled from 5 different datasets) with normal anatomy, including a variety of image qualities with most images defaced. The data also include detailed brain parcellation with 108 classes. We used the smallest dataset ('Extra_18_volumes') to identify 77 classes that unambiguously corresponded to gray and white matter within cerebrum, cerebellum, and brainstem. Subcortical and midbrain areas were excluded from Dice score computation (45 classes) on the rest of the data (N=83 subjects).

**2.9** *Effects of segmentation and electrode placement on current-flow models*

We also evaluate the effect of different segmentation on current-flow models using the software ROAST. For this we calculated the lead field (using the 'leadField' option to the roast command) with Multipriors, Multiaxial, and manual ground truth labels. Current-flow models also depend crucially on the placement of electrodes on the segmented head. We tested automatic placement in ROAST, which leverages SPM12 alignment, and manual placement. Next, we selected an optimal electrode montage (using the roast_target[48–50] command), to target a location near the lesion that corresponds to gray matter in all segmentations, and computed the resulting electric field distributions across the brain. We repeated modeling and targeting on 10 aphasia stroke cases that were previously identified as misaligned. For each case, we computed results at 5 different target locations and evaluated the electric field at those targets.

## 3  RESULTS

We utilized the validation set performance to select hyperparameters and model architecture (Fig. 2, Table S1). Based on this, we train a model with a depth of 6 layers (1,016,839 trainable parameters, Fig. 2C) and data augmentation consisting of random rotations and shearings at 15 degrees and addition of spatial information as an extra input (Fig. 2A). We add an extra input with voxel coordinates such that each model has global spatial information (Fig. 2B). We also compare two ways of merging the outputs of the models operating in each axis (sagittal, coronal and axial) by implementing a majority vote method as well as a convolutional layer (Fig. 2D).

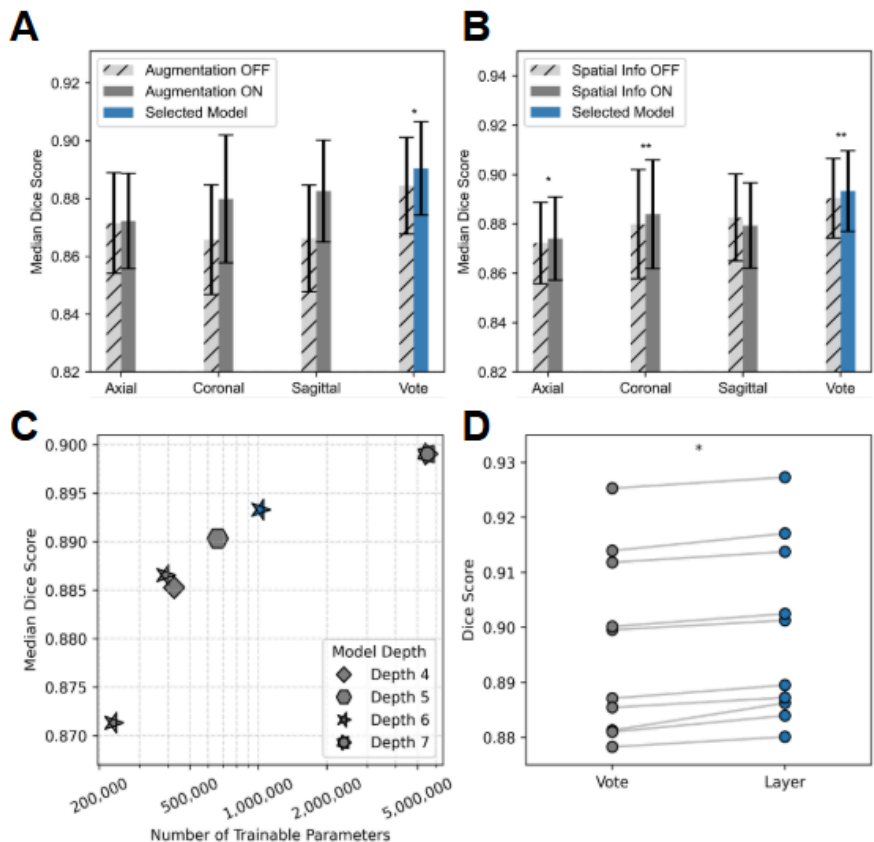

**Figure 2: Model selection on a validation set.** A: Determines the benefit on Dice score of including data augmentation. B: Determines the benefit on Dice score of including voxel coordinates as additional input. C: Determines the benefit of increasing depth and width of the U-Net, which implies increase in the number of parameters. D: Determines the benefit of using a CNN consensus layer compared with a median vote consensus for merging the output of the three models operating in each orthogonal direction. Significant pairwise differences are indicated as corrected *p< 0.05,**p< 0.01,***p < 0.001.

**3.1** *Performance comparison with other segmentation tools for head MRI*

We compare the performance of the MultiAxial model on our test set (Table 1) against the MultiPriors model (which depends on SPM for alignment with a TPM), as well as SPM12's segmentation tool. Dice scores were significantly higher for the MultiAxial model as compared to the other models (Fig. 3A, median, mean, and statistical results in Table S2). Noticeably, SPM12 struggles more often in the TPM alignment step, influencing the resulting segmentation (Fig. 3C). Dice scores for brain tissue were compared against two publicly available AI-tools SynthSeg and Brainchop, as these models only provide labels for brain structures. MultiAxial significantly outperformed both models in our test set of Table 1 (Fig. 3B, 3D, median, mean, and statistical results in Table S3). This result is not surprising given that these models produce parcellations that do not unambiguously map to gray and white matter in our manual ground truth (see caveats).

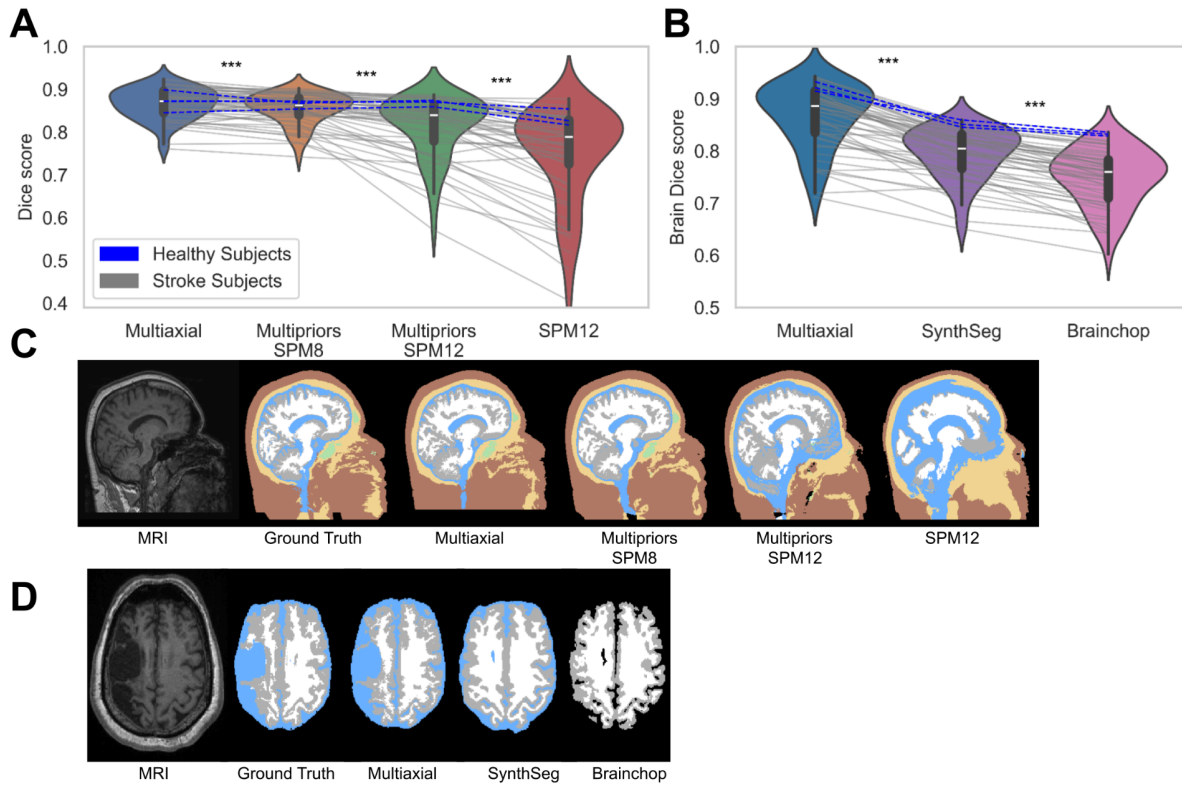

**Figure 3: Performance comparison between segmentation tools.** A**:** Test set Dice scores for head MRI segmentation with Multiaxial and Multipriors. The alignment of the atlas required by Multipriors was done with ROAST, which relies on SPM8 or SPM12 depending on version number**.** B: Test set Dice scores for segmentation of brain (white matter and gray matter) using Multiaxial, SynthSeg and Brainchop. C: Example segmentation (sagittal view) for models compared in panel A, in the same order. D: Example brain segmentation (axial view) for models

compared in panel B, in the same order. The ground-truth shows a CSF-filled stroke lesion in the left hemisphere that is missed by SynthSeg and Brainchop. (*p<0.05,**p<0.01,***p<0.001)

To address this caveat, we test the models also on an out-of-sample dataset where the ground truth labels are brain parcellations (Mindboggle). This data includes only normal brain anatomy with 4 distinct subsets of data. SynthSeg outperforms the other methods (median, mean, and statistical results in Table S4). Dice scores are in a similar range for the NKI datasets (Fig. 4A) and provide comparable segmentation (Fig. 4C) but differs substantially for the other two (Fig. 4B), because they deviate substantially in appearance from our data sets (see Discussion).

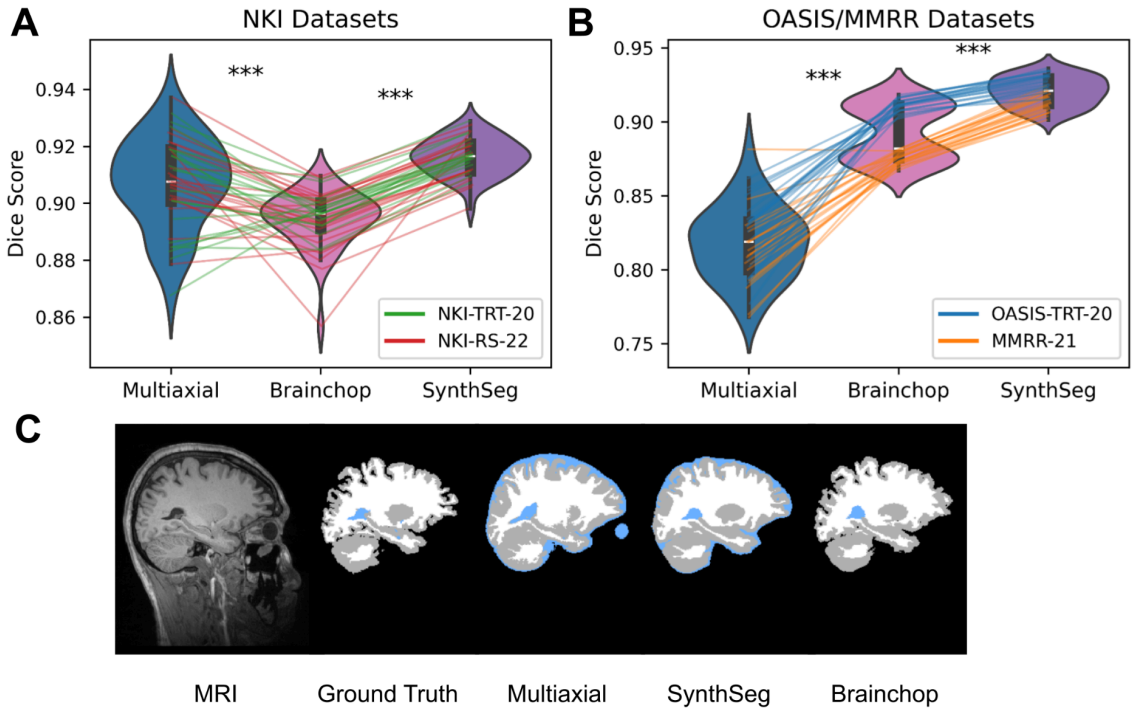

**Figure 4: Out-of-sample evaluation.** Dice score for brain segmentation on Mindboggle data only including parcels that exclusively cover white or gray matter. A: NKI subset have MRI of the full head in with conventional T1 images. B: The OASIS dataset is missing much of the head anatomy. The MMRR dataset has darker gray matter than usual T1 images. C: Example segmentation (sagittal view) for models compared in panel A.

**3.2** *Anonymization of head MRI and model performance in defaced data*

The MultiAxial model achieved similar Dice scores for defaced and refaced MRI images (Fig. 5 median, mean, and statistical results in Table S5). This was evaluated on structures inside the skull, which should remain unaltered by defacing, including gray matter and white matter. The 3 types of MRI's had similar numerical performance, this demonstrates that, even when using anonymized MRIs such as defaced and refaced data, the model maintains strong performance with comparable Dice scores despite significant differences between groups.

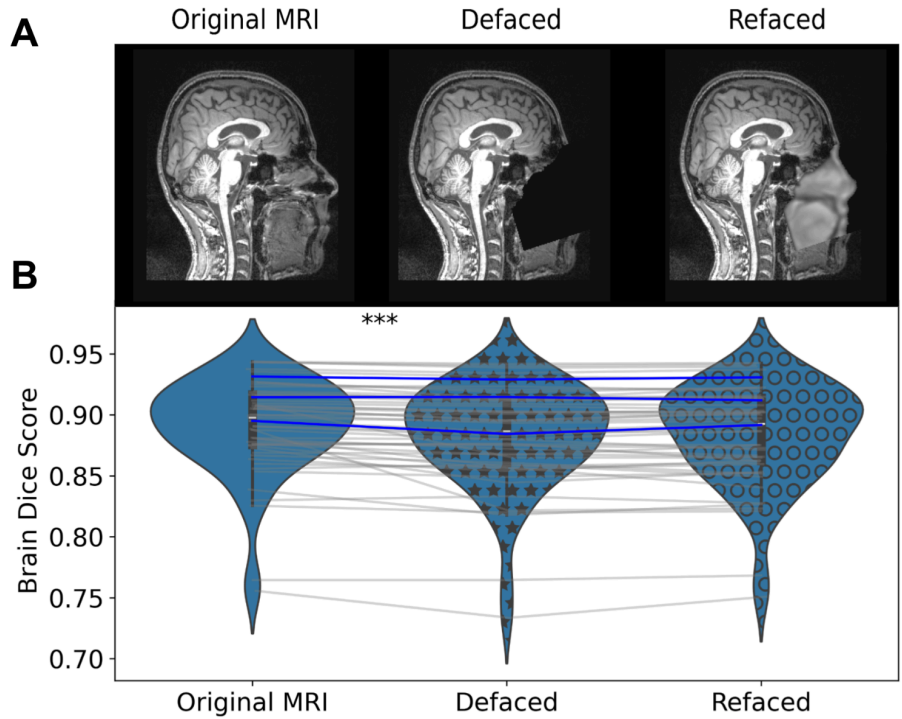

**Figure 5: Robust segmentation of the brain on de-identified head MRI.** A: Example of two de-identification methods in ANFI applied to a healthy subject: 'Deface' removes identifiable features by removing the face from the image. 'Reface' replaces the face with an average of multiple aligned faces. B: Dice scores for segmentation of brain tissue (white matter and gray matter) on the original MRI, defaced MRI and 'Refaced' MRI. (corrected *p< 0.05,**p< 0.01,***p < 0.001)

### 3.3 *Improvement of a current flow modelling software*

To test the effect of segmentations on current-flow models we used the ROAST software. In order to inspect the sensitivity of the field magnitude to segmentations we used the Multiaxial and Multipriors models on 10 stroke heads. An example of the field distribution for a particular target location in the brain is shown in (Fig. 6A). For the practical use of current field modeling it is important that electrode locations are correctly placed on standard locations on the scalp (Fig. S2). We tested manual electrode placement as well as automatic placement, that relies on SPM12. We compared the electric field magnitude on the target location obtained with the ground truth segmentation. The results favour the Multiaxial model with manual electrode placement (Fig. 6B, median, mean, and statistical results in Table S6).

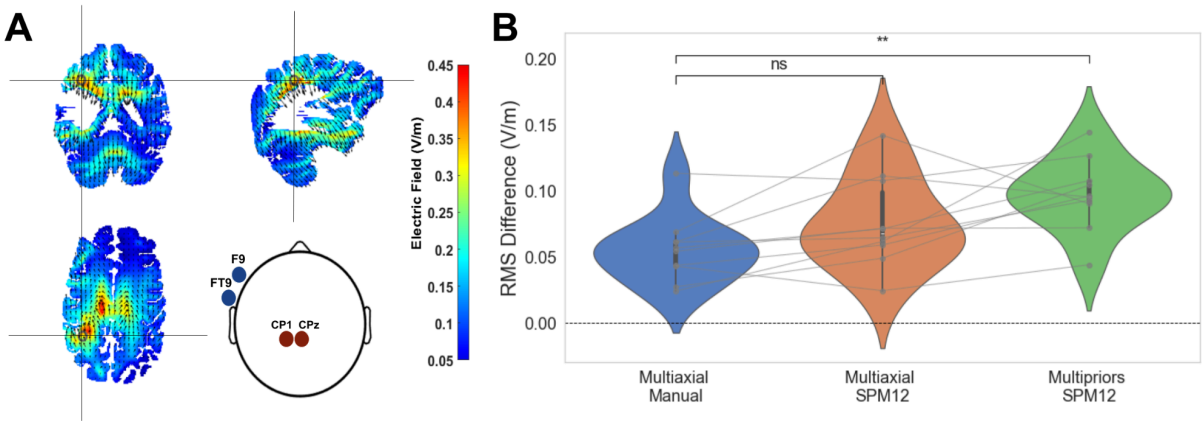

**Figure 6: Current flow modelling in the brain given a target.** An MRI from a single subject with a large stroke lesion was used for current flow modeling using the software 'ROAST'. A: Simulated electric field in the brain for a particular electrode montage (bottom right). This montage was optimized to generate maximum field magnitude at the target location (cross hair) under maximum 1mA current in each electrode. This simulation used the Ground Truth segmentation. Comparison of the electric field magnitude at the specified target when using different methods for segmentation. B: Difference in the electric field magnitude from those obtained with ground truth segmentation and manual electrode placement. Distance is measured as root mean square (RMS) averaged over 5 different target locations (corrected *p< 0.05,**p< 0.01,***p < 0.001)

### 3.4 *Benefit to processing speed*

An advantage of MultiAxial is processing speed. As the model operates in 2D it requires less memory and gains in speed compared to previous methods (see Table 2), with exception of Brainchop that has been optimized for GPU usage.

**Table 2:** Processing speeds in minutes:seconds evaluated on the same MRI either using a GPU (NVIDIA RTX A5000, 24 GB RAM) or a CPU (Intel Core i7-8700K, 64 GB RAM) on Windows 11.

|            | MultiAxial | MultiPriors* | SPM8 | SPM12 | Brainchop | SynthSeg |
|------------|------------|--------------|------|-------|-----------|----------|
| CPU time   | 1:19       | 2:23         | 9:49 | 7:09  | 17:29     | 7:46     |
| GPU time   | 0:20       | 1:18         | NA   | NA    | 0:08      | 1:28     |
| # of param | 3.25M      | 0.75M        | NA   | NA    | 0.77M     | 52.5M    |

* Includes warping for spatial alignment of a TPM, but assumes that the warping field has been pre-computed with SPM. All other times include all required inference, pre- and post-processing steps.

## 4 DISCUSSION

The most wide-spread tools for brain signal analysis require a detailed segmentation of the brain, and to do so, they rely on an atlas for segmentation, such as a TPM in SPM, or in FAST as part of FSL[51], or a multi-atlas segmentation in ANTs[52]. These methods are susceptible to misalignment when registering an atlas of normal anatomy to an individual MRI with abnormal anatomy. Additionally, these atlases assume normal anatomy and thus introduce incorrect biases in the segmentation. The new MultiAxial model eliminates the dependency on an atlas, thereby simplifying the processing and gaining in robustness. We overcame this dependency by training the AI on a large number of heads with abnormal anatomy (N=76 for training plus N=12 for validation). The resulting model shows state-of-the-art performance on whole head segmentation with abnormal brain anatomies. Crucially, we are publicly releasing the training data and model to serve as a benchmark and basis for future development efforts.

The dataset compiled and released here is unique in that we included cases with chronic stroke and traumatic brain injury. These cases are characterized by extended ventricles and fluid-filled voids where one would usually expect brain tissue. In contrast, existing clinical datasets are mostly focused on brain tumors, such as the International Brain Tumor Segmentation (BraTS) Challenge, which provides labels only for the tumors. The 2012 dataset contained 65 high and low grade gliomas[53]. The 2015 dataset contains 220 high grade and 54 low grade gliomas for training and 94 unpublished cases reserved for blinded testing[54]. This dataset was recently extended in the BraTS 2024 release with approximately 4,500 cases.[55,56].

Our focus in compiling new data was unique in that we generated labels and evaluated performance on the entire head, not just the brain. There is in general a shortage of high-quality manual annotated MRI datasets.[57,58] Comprehensive manual corrections and whole-head manual tissue annotations in 3D are particularly time-consuming and rely on trained personnel. The majority of public brain MRI datasets are skull-stripped (BraTS[55], MICA[59], HCP[60], Mindboggle[46], ADNI[61]), omitting key anatomical features outside the brain, which restricts their utility for tasks requiring full-head analysis, such as modeling of current flow in brain stimulation studies. As a result, datasets that have been manually segmented are often small in size, such as those used for testing of SynthSeg [32] (N=39) or Brainchop [31] (N=5). Another factor to consider is the large size of our dataset. When evaluating manually labeled pixels excluding background, our dataset stands as the largest, surpassing even well-known large-scale datasets like BraTS[55] and Mindboggle[46] (Table 6). Despite having fewer subjects, our dataset contains significantly more labeled data, highlighting the density and detail of our annotations.

**Table 3:** Comparison of non-background labeled pixels with popular head MRI datasets.

| Data set | N = Subjects | N = Labeled Pixels | Labeled Pixels/Subject |
|---|---|---|---|
| BraTS | 1621 | $119 \times 10^6$ | $73.3 \times 10^3$ |
| Mindboggle | 101 | $47 \times 10^6$ | $463 \times 10^3$ |
| This work | 98 | $461 \times 10^6$ | $4.70 \times 10^6$ |

Newer AI models, such as Brainchop and SynthSeg are trained on large datasets segmented with FreeSurfer[62] to generate brain parcellations including gray and white matter as well as subcortical structures. Brainchop,[31] was designed to make brain parcellation easily accessible over web browsers, and employs the Catalyst.Neuro's model [47,63,64]. It used 1200 healthy heads from Human Connectome Project (HCP)[60] and 101 heads from Mindboggle [46,65]. SynthSeg [32,66] was developed to generate parcelations on any kind of brain images including CT and was trained using automated segmentation labels from 686 heads including from healthy subjects and patients with memory complaints and Alzheimer's from Alzheimer's Disease Neuroimaging Initiative (ADNI)[61]. It used these labels to generate synthetic images for training, hence the name SynthSeg. The focus of these models is parcellation of the brain, and assume that the overall brain anatomy is well captured with traditional segmentation methods. However, as we have seen, these models struggle with abnormalities that erroneously classify stroke damage as gray or white matter when the area is now filled with CSF. The present model, to our knowledge, is the only model to have been trained on chronic stroke and disorder of consciousness patients, that are characterized by large CSF-filled voids.

Many older segmentation tools suffer from outdated codebases, poor documentation, and reliance on deprecated Python libraries. In contrast, our model is fully integrated into ROAST, offering ease of use. Its batch script automatically sets up the environment, enabling segmentation without requiring command-line expertise. Furthermore, our model is lightweight, self-contained and boasts minimal dependencies, ensuring seamless compatibility across Windows, macOS, and Linux—a clear advantage over older tools that often exclude Windows compatibility such as SynthSeg[32], AFNI [52], FreeSurfer[62], FSL[16], Deepmedic[17] which require implementing a Linux or macOS environment. Such requirements can present a considerable barrier for new users, making our solution a more accessible and user-friendly alternative.

The MultiAxial model has a field of view (FOV) equal to a complete side-view of the MRI in each direction (sagittal, axial and coronal), meaning that the prediction output for any given pixel is influenced by the input features from locations far away. This is in contrast to common implementations of 3D networks (such as the DeepMedic,[17] of MultiPriors[25] or nnUNET[67]) which, in order to satisfy VRAM limitations, make predictions based on small cubes at a time. A larger FOV provides spatial awareness which reduces mistakes, which is why methods such as SPM and MultiPriors rely on tissue probability maps for a proper result. On the other hand we observed that distant regions influence the model outputs: Defaced inputs have slightly different segmentation results on the brain, despite the defacing process not affecting the brain areas. Additionally, the 2D processing may lead to discontinuities in the off-plane directions, less likely to occur in 3D networks. However, we find that the 3D consensus layer smooths our discontinuities.

Our approach to segmentation is similar to previous efforts to brain segmentation in the presence of lesions. One study used a large population of diverse clinical scans to train a 3D UNet[68], not unlike our MultiPriors model. As with that model, it relied on alignment with a template. It is this dependence on registration that proved problematic and motivated us to develop a more robust tool. An earlier study focusing on Multiple Sclerosis lesions used a 2D UNet in axial direction, comparable to our axial network [69]. We showed here that this approach can be improved by training additional networks in sagittal and coronal planes. To combine these planes our model employs a consensus layer rather than traditional ensemble methods like voting or averaging. Voting combines 2D segmentations by selecting the pixel class with mutual agreement. In cases of disagreement, it defaults to the most common or highest-confidence class. Averaging is typically used for binary classifications like lesions vs. non-lesions but is unsuitable for our multi-class segmentation. Our 3D consensus layer integrates segmentations from all three 2D views and the MRI data to achieve superior segmentation accuracy.

In this work, the Dice score for the brain is usually around 90% and a few points less when including the entire head. What is the source of the remaining 10% errors? We have shown evidence that some of these errors may be due to inaccurate "truth data". As in previous work, the target labels were largely generated automatically. We have invested a significant effort with manually correcting these segmentations on all the data we are releasing. Nevertheless, we believe that the network may produce in some places more accurate segmentations than the manually corrected labels. This is consistent with the somewhat higher performance for the brain, as manual correction was generally more careful for the brain and CSF in this data, and less effort went into fixing the non-brain soft tissue covering skin, scalp and other soft tissues, which all received a single label. On the flip side, we acknowledge an upwards bias in the results, because SPM and MultiPriors were used to produce the first round of "truth" labels, and the network was trained on those. The MultiAxial network was trained to reproduce the result of those algorithms and may thus have an upwards bias in performance. In summary, the 10% residual "error" in performance may be an optimistic estimate due to the bias of semi automatic truth labels, or it may be an underestimate due to insufficient manual corrections in the truth data.

On normal anatomy we achieve a median performance of 0.915, which is comparable to FreeSurfer[70] which reports Dice scores around 0.91[47] on gray and white matter segmentation. Here we found approximately 0.91-0.92 for both Multiaxial and SynthSeg on out-of-sample data (NKI as part of Mindboggle), but poor performance for Multiaxial when the MRI differed significantly in appearance from our training data. In the MMRR dataset gray matter is with CSF because it is particularly dark, likely due to an unconventional T1 imaging sequence used in that study[71]. In the OASIS dataset the de-identification removed not just the face but jaw, ears, neck and spine. Performance on non-brain classes is rarely reported in the literature, but in the present data we significantly outperformed Multipriors and SPM12. In total, the current model, Multiaxial, sets a new state-of-the-art for non-brain tissue, and outperformed existing tools on gray and white matter in the presence of abnormal brain anatomy. However, more work is needed to make the MultiAxial model robust to different forms of de-identified MRI images and more diverse imaging sequences.

A limitation of our method is that it does not produce the detailed parcelations generated by typical brain segmentation algorithms. Future work may be able to extend our approach by retraining the mode with automated labels that merge, say FSL parcelations, with CSF, skull, skin labels with our existing tool. An additional limitation is that our training data mostly focused on chronic stroke and with disorders of consciousness who typically present with abnormal anatomies. Future work will need to explore if the trained model generalizes to cases that include tumors, multiple sclerosis lesions, hydrocephalus or congenital malformations. As with parcelations, such work may benefit from retraining with merged labels leveraging existing partially labeled brain MRIs in those populations, e.g. BraTS [55,56].

As a caveat, we should add that the multi-planar modeling approach used here is not new, having been used previously for 3D MRI [27–30], although not specifically in the context of abnormal head anatomies. We see part of the appeal of the MultiAxial model in its simplicity that allows for easy training. As a tool, ready-to-use for those users needing segmentation of the whole head, it is the best available model. For users interested in brain parcellation, other trained models such as SynthSeg or BrainChop will be better suited.

A caveat to the comparison with SynthSeg and Brainchop is that these tools are trained to produce parcelations of the brain, while our labels demarcate gray and white matter. Most parcels map uniquely to gray or white matter, but that is not the case for some subcortical regions, e.g. the basal ganglia contains both white and gray matter (see Fig. S3). Therefore, there is a systematic bias favouring the Multiaxial model when evaluating performance on our gray/white matter labels (Fig. 3B). When evaluating performance on the Mindboggle data we can exclude those parcelations, and since neither SynthSeg nor MultiAxial were trained on these labels, the resulting Dice scores (Fig. 3E) represents an unbiased estimate of performance. Despite the bias, the results are meaningful for those users that require a tool for gray/white matter segmentation. These unbiased results may be of interest for developers that intend to train models on new data.

To support further research, we are releasing a manually labeled stroke MRI dataset unavailable elsewhere. This dataset includes annotations for seven tissue classes, offering invaluable resources for advancements in stroke research and related applications.

## 5 Conclusion

We are publicly releasing a ready-to-use, state-of-art tool for whole-head MRI segmentation which is robust to clinical abnormalities, as well as a labeled data-set of clinical MRIs. We also integrated the model into ROAST, an existing pipeline for current flow simulation with an easy-to-use user-interface, making it robust for abnormal clinical MRIs where previous software failed.

## 6 Acknowledgments

We would like to thank Jacobo D. Sitt at Pitie-Salpetriere University Hospital in Paris for the help in gathering and sharing the dataset on disorders of consciousness (not released). The work was supported by NIH grants R01DC018589, R01NS130484, R01CA247910, R44NS092144.

## 7 Disclosures

The authors declare that there are no financial interests, commercial affiliations, or other potential conflicts of interest that could have influenced the objectivity of this research or the writing of this paper.

## 8 Code and Data Availability

The model was implemented using Tensorflow and the code is available on Github at https://github.com/lkshrsch/multiaxial_brain_segmenter. The data summarized in Table 1 includes for each subject a T1 image with isotropic resolution in RSA orientation, and segmentation labels for seven categories and a T1 image for each subject. It is available on Kaggle at https://www.kaggle.com/datasets/andrewbirnbaum/full-head-mri-and-segmentation-of-stroke-patients. A notebook demonstrating the use of the model and data is also on the same Kaggle repository: https://www.kaggle.com/code/andrewbirnbaum/multiaxial-demo.

**Biographies**

Andrew M. Birnbaum earned his Master's degree in Biomedical Engineering from the City College of New York in 2025, graduating as valedictorian. At the Parra Lab, he focuses on applications of transcranial direct current stimulation (tDCS), developing software to model current flow in abnormal brain anatomy. He has helped adapt the ROAST pipeline for atypical anatomies and contributed to preclinical research demonstrating the effectiveness of tDCS in human and rodent models.

Adam Buchwald is a Professor and principal investigator at New York University in the department of Communicative Sciences and Disorders. His research focuses on speech and language production and neurorehabilitation following stroke. His research includes mechanistic studies of spoken production as well as rehabilitation science studies combining speech motor learning treatment and neuromodulation to enhance stroke outcomes.

Peter Turkeltaub is a cognitive neurologist and neuroscientist who specializes in aphasia neurorehabilitation and the brain basis of language. He earned an MD and PhD in neuroscience from Georgetown University and completed his neurology residency and cognitive neurology fellowship at the University of Pennsylvania. Dr. Turkeltaub directs the Cognitive Neurology Lab and the Aphasia Clinic at MedStar National Rehabilitation Hospital.

Adam Jacks completed his doctorate in communication sciences and disorders at The University of Texas at Austin and postdoctoral training in behavioral neuroimaging in San Antonio. His research explores how neuropathology affects neuromotor speech disorders using methods such as fMRI, acoustic/perceptual speech analysis, and speech perturbation. By studying motor control in typical and disordered speech, Dr. Jacks aims to improve prediction of speech outcomes and guide more effective treatments for individuals with neurological injuries or motor speech disorders.

Yu Huang, Ph.D. is a Senior Research Scientist at Soterix Medical Inc.. He is trained as a biomedical engineer. He specializes in large-scale medical image processing, neural signal processing, biophysical modeling and optimization, statistical machine learning, deep learning, and their clinical applications. He created a pipeline software named ROAST for processing human head MRIs and building computational models for predicting electric current flow in the head.

Abhishek Datta, PhD is the head of R&D at Soterix Medical. He is also an Adjunct Professor of Biomedical Engineering at City College of New York. His research interests include developing non-invasive brain stimulation technology, computational modeling and validation, and optimization of brain stimulation modalities.

George Carr is a neural engineering intern in the Parra Lab, where he works on tissue segmentation, anatomical labeling, medical imaging, and exploring clinical applications of language models. He transitioned to neuroscience after a career in the arts and holds an MFA from Yale University. He earned a BS in Psychology from the City College of New York in 2024. His research interests include neural repair, functional neurosurgery, and the development of therapeutic neurotechnology. He plans to attend medical school.

Shreya Kannan is currently an undergraduate student in biomedical engineering at the City College of New York. Her research currently includes work in robotic surgical technologies, neural imaging, and cognitive evaluations within the context of neurodegenerative disorders. Her industry experience lies in testing and R&D at Medtronic Surgical Robotics.

Lucas C. Parra is the Harold Shames Professor of Biomedical Engineering at the City College of New York. He received his Ph.D. in Physics from Ludwig-Maximilians-Universität in 1996, focusing on machine learning. His early career included work on medical imaging at Siemens Corporate Research (1995-1997) and acoustic signal processing at Sarnoff Corporation (1997-2003). His research interests include neural signals, brain stimulation, and machine learning, with a focus on the brain's responses to natural stimuli.

Lukas Hirsch earned his Ph.D. in Biomedical Engineering from City College of New York in September 2023. As a researcher in the Parra Lab, his work focuses on computer vision and AI applications in medical imaging, particularly for the early detection of breast cancer using MRI. He has contributed to studies demonstrating AI's capability to identify breast cancer significantly earlier than traditional methods, aiming to improve screening outcomes for high-risk women. He currently works at Janssen R&D.

**Figure Captions**

**Figure 1: Overview of the MultiAxial model for segmentation of head MRI.** An input T1 weighted MRI is processed in each axis (axial, sagittal and coronal) by a separate 2D U-Net, each of which output probabilities per pixel for belonging to one of seven classes: Background (black), Air/Sinus cavities (light green), White Matter (white), Gray Matter (gray), Cerebrospinal fluid (CSF, light blue), bone (light yellow) and skin (brown). The probabilities from each model are merged into a single volume by the consensus layer, consisting of a single 3D convolutional layer.

**Figure 2: Model selection on a validation set.** A: Determines the benefit on Dice score of including data augmentation. B: Determines the benefit on Dice score of including voxel coordinates as additional input. C: Determines the benefit of increasing depth and width of the U-Net, which implies increase in the number of parameters. D: Determines the benefit of using a CNN consensus layer compared with a median vote consensus for merging the output of the three models operating in each orthogonal direction. Significant pairwise differences are indicated as corrected *$p< 0.05$,**$p< 0.01$,***$p < 0.001$.

**Figure 3: Performance comparison between segmentation tools.** A: Test set Dice scores for head MRI segmentation with Multiaxial and Multipriors. The alignment of the atlas required by Multipriors was done with ROAST, which relies on SPM8 or SPM12 depending on version number. B: Test set Dice scores for segmentation of brain (white matter and gray matter) using Multiaxial, SynthSeg and Brainchop. C: Example segmentation (sagittal view) for models compared in panel A, in the same order. D: Example brain segmentation (axial view) for models compared in panel B, in the same order. The ground-truth shows a CSF-filled stroke lesion in the left hemisphere that is missed by SynthSeg and Brainchop. (*$p<0.05$,**$p<0.01$,***$p<0.001$)

**Figure 4: Out-of-sample evaluation.** Dice score for brain segmentation on Mindboggle data only including parcels that exclusively cover white or gray matter. A: NKI subset have MRI of the full head in with conventional T1 images. B: The OASIS dataset is missing much of the head anatomy. The MMRR dataset has darker gray matter than usual T1 images. C: Example segmentation (sagittal view) for models compared in panel A.

**Figure 5: Robust segmentation of the brain on de-identified head MRI.** A: Example of two de-identification methods in ANFI applied to a healthy subject: 'Deface' removes identifiable features by removing the face from the image. 'Reface' replaces the face with an average of multiple aligned faces. B: Dice scores for segmentation of brain tissue (white matter and gray matter) on the original MRI, defaced MRI and 'Refaced' MRI. (corrected *$p< 0.05$,**$p< 0.01$,***$p < 0.001$)

**Figure 6: Current flow modelling in the brain given a target.** An MRI from a single subject with a large stroke lesion was used for current flow modeling using the software 'ROAST'. A: Simulated electric field in the brain for a particular electrode montage (bottom right). This montage was optimized to generate maximum field magnitude at the target location (cross hair) under maximum 1mA current in each electrode. This simulation used the Ground Truth segmentation. Comparison of the electric field magnitude at the specified target when using different methods for segmentation. B: Difference in the electric field magnitude from those obtained with ground truth segmentation and manual electrode placement. Distance is measured as root mean square (RMS) averaged over 5 different target locations (corrected *$p< 0.05$,**$p< 0.01$,***$p < 0.001$)

## References

1. Stephen J. Gardner MS, Joshua Kim PhD, and Indrin J. Chetty PhD, "Modern Radiation Therapy Planning and Delivery," Hematol. Oncol. Clin. North Am. **33**(6), 947–962 (2019) [doi:https://doi.org/10.1016/j.hoc.2019.08.006.].
2. H. Hyare, S. Thust, and J. Rees, "Advanced MRI Techniques in the Monitoring of Treatment of Gliomas," Curr. Treat. Options Neurol. **19**(3), 11 (2017) [doi:10.1007/s11940-017-0445-6].
3. H. Allioui, M. Sadgal, and A. Elfazziki, "Deep MRI Segmentation: A Convolutional Method Applied to Alzheimer Disease Detection," Int. J. Adv. Comput. Sci. Appl. **10**(11) (2019) [doi:10.14569/IJACSA.2019.0101151].
4. C. Ledig et al., "Robust whole-brain segmentation: Application to traumatic brain injury," Med. Image Anal. **21**(1), 40–58 (2015) [doi:10.1016/j.media.2014.12.003].
5. X. Ren et al., "Robust Brain Magnetic Resonance Image Segmentation for Hydrocephalus Patients: Hard and Soft Attention," in 2020 IEEE 17th International Symposium on Biomedical Imaging (ISBI), pp. 385–389 (2020) [doi:10.1109/ISBI45749.2020.9098541].
6. A. Clèrigues et al., "Acute and sub-acute stroke lesion segmentation from multimodal MRI," Comput. Methods Programs Biomed. **194**, 105521 (2020) [doi:10.1016/j.cmpb.2020.105521].
7. J. A. Stokum, V. Gerzanich, and J. M. Simard, "Molecular pathophysiology of cerebral edema," J. Cereb. Blood Flow Metab. **36**(3), 513–538, SAGE Publications Ltd STM (2016) [doi:10.1177/0271678X15617172].
8. Juhasz C et al., "White matter volume as a major predictor of cognitive function in Sturge-Weber syndrome," Arch. Neurol. **64**(8), 1169–1174, American Medical Assn, United States (2007) [doi:10.1001/archneur.64.8.1169].
9. J. A. Meltzer et al., "Individual differences in EEG theta and alpha dynamics during working memory correlate with fMRI responses across subjects," Clin. Neurophysiol. Off. J. Int. Fed. Clin. Neurophysiol. **118**(11), 2419–2436 (2007) [doi:10.1016/j.clinph.2007.07.023].
10. A. Datta, "Inter-Individual Variation during Transcranial Direct Current Stimulation and Normalization of Dose Using MRI-Derived Computational Models," Front. Psychiatry **3**, Frontiers (2012) [doi:10.3389/fpsyt.2012.00091].
11. Y. Huang et al., "ROAST: An Open-Source, Fully-Automated, Realistic Volumetric-Approach-Based Simulator For TES," in 2018 40th Annual International Conference of the IEEE Engineering in Medicine and Biology Society (EMBC), pp. 3072–3075 (2018) [doi:10.1109/EMBC.2018.8513086].
12. Y. Huang et al., "Realistic volumetric-approach to simulate transcranial electric stimulation-ROAST-a fully automated open-source pipeline," J. Neural Eng. **16**(5), 056006 (2019) [doi:10.1088/1741-2552/ab208d].
13. F. Tadel et al., "Brainstorm: A User-Friendly Application for MEG/EEG Analysis," Comput. Intell. Neurosci. **2011**, 879716 (2011) [doi:10.1155/2011/879716].
14. A. Thielscher, A. Antunes, and G. B. Saturnino, "Field modeling for transcranial magnetic stimulation: A useful tool to understand the physiological effects of TMS?," in 2015 37th Annual International Conference of the IEEE Engineering in Medicine and Biology Society (EMBC), pp. 222–225 (2015) [doi:10.1109/EMBC.2015.7318340].
15. J. Ashburner and K. J. Friston, "Unified segmentation," NeuroImage **26**(3), 839–851 (2005) [doi:10.1016/j.neuroimage.2005.02.018].
16. S. M. Smith et al., "Advances in functional and structural MR image analysis and implementation as FSL," NeuroImage **23 Suppl 1**, S208-219 (2004) [doi:10.1016/j.neuroimage.2004.07.051].
17. K. Kamnitsas et al., "DeepMedic for Brain Tumor Segmentation," 2016, Cham, 138–149, Springer International Publishing [doi:10.1007/978-3-319-55524-9_14].
18. S. Nigro et al., "Fully Automated Segmentation of the Pons and Midbrain Using Human T1

MR Brain Images," PLOS ONE **9**(1), e85618, Public Library of Science (2014) [doi:10.1371/journal.pone.0085618].
19. K. Kushibar et al., "Automated sub-cortical brain structure segmentation combining spatial and deep convolutional features," Med. Image Anal. **48**, 177–186 (2018) [doi:10.1016/j.media.2018.06.006].
20. S. Pereira et al., "Brain Tumor Segmentation Using Convolutional Neural Networks in MRI Images," IEEE Trans. Med. Imaging **35**(5), 1240–1251 (2016) [doi:10.1109/TMI.2016.2538465].
21. C. Gros et al., "Automatic segmentation of the spinal cord and intramedullary multiple sclerosis lesions with convolutional neural networks," NeuroImage **184**, 901–915 (2019) [doi:10.1016/j.neuroimage.2018.09.081].
22. H. K. van der Burgh et al., "Deep learning predictions of survival based on MRI in amyotrophic lateral sclerosis," NeuroImage Clin. **13**, 361–369 (2017) [doi:10.1016/j.nicl.2016.10.008].
23. D. Pustina et al., "Automated segmentation of chronic stroke lesions using LINDA: Lesion identification with neighborhood data analysis," Hum. Brain Mapp. **37**(Pustina, Dorian, H. Branch Coslett, Peter E. Turkeltaub, Nicholas Tustison, Myrna F. Schwartz, and Brian Avants. "Automated Segmentation of Chronic Stroke Lesions Using LINDA: Lesion Identification with Neighborhood Data Analysis." Human Brain Mapping 37, 4 (2016): 1405–21. https://doi.org/10.1002/hbm.23110.), 1405–1421 (2016) [doi:10.1002/hbm.23110].
24. H. McCann, G. Pisano, and L. Beltrachini, "Variation in Reported Human Head Tissue Electrical Conductivity Values," Brain Topogr. **32**(5), 825–858 (2019) [doi:10.1007/s10548-019-00710-2].
25. L. Hirsch, Y. Huang, and L. C. Parra, "Segmentation of MRI head anatomy using deep volumetric networks and multiple spatial priors," J. Med. Imaging Bellingham Wash **8**(3), 034001 (2021) [doi:10.1117/1.JMI.8.3.034001].
26. O. Ronneberger, P. Fischer, and T. Brox, "U-Net: Convolutional Networks for Biomedical Image Segmentation," in Medical Image Computing and Computer-Assisted Intervention – MICCAI 2015, N. Navab et al., Eds., pp. 234–241, Springer International Publishing, Cham (2015) [doi:10.1007/978-3-319-24574-4_28].
27. G. Piantadosi et al., "Multi-planar 3D breast segmentation in MRI via deep convolutional neural networks," Artif. Intell. Med. **103**, 101781 (2020) [doi:10.1016/j.artmed.2019.101781].
28. M. F. Stollenga et al., "Parallel multi-dimensional LSTM, with application to fast biomedical volumetric image segmentation," in Proceedings of the 29th International Conference on Neural Information Processing Systems - Volume 2 **2**, pp. 2998–3006, MIT Press, Cambridge, MA, USA (2015).
29. H. Wang et al., "Mixed 2D and 3D convolutional network with multi-scale context for lesion segmentation in breast DCE-MRI," Biomed. Signal Process. Control **68**, 102607 (2021) [doi:10.1016/j.bspc.2021.102607].
30. S. Aslani et al., "Deep 2D Encoder-Decoder Convolutional Neural Network for Multiple Sclerosis Lesion Segmentation in Brain MRI," in Brainlesion: Glioma, Multiple Sclerosis, Stroke and Traumatic Brain Injuries, A. Crimi et al., Eds., pp. 132–141, Springer International Publishing, Cham (2019) [doi:10.1007/978-3-030-11723-8_13].
31. M. Masoud, F. Hu, and S. Plis, "Brainchop: In-browser MRI volumetric segmentation and rendering," J. Open Source Softw. **8**(83), 5098 (2023) [doi:10.21105/joss.05098].
32. B. Billot et al., "SynthSeg: Segmentation of brain MRI scans of any contrast and resolution without retraining," Med. Image Anal. **86**, 102789 (2023) [doi:10.1016/j.media.2023.102789].
33. A. Kirillov et al., "Segment Anything," presented at Proceedings of the IEEE/CVF International Conference on Computer Vision, 2023, 4015–4026.
34. J. Ma et al., "Segment anything in medical images," Nat. Commun. **15**(1), 654, Nature Publishing Group (2024) [doi:10.1038/s41467-024-44824-z].

35. Y. Huang et al., "Automated MRI Segmentation for Individualized Modeling of Current Flow in the Human Head," J. Neural Eng. **10**(6), 10.1088/1741-2560/10/6/066004 (2013) [doi:10.1088/1741-2560/10/6/066004].
36. A. Datta et al., "Gyri-precise head model of transcranial direct current stimulation: Improved spatial focality using a ring electrode versus conventional rectangular pad," Brain Stimulat. **2**(4), 201-207.e1 (2009) [doi:10.1016/j.brs.2009.03.005].
37. C. Rorden and M. Brett, "Stereotaxic display of brain lesions," Behav. Neurol. **12**(4), 191–200 (2000) [doi:10.1155/2000/421719].
38. R. Cox and P. Taylor, "Why de-face when you can re-face?," 26th Annual Meeting of the Organization for Human Brain Mapping (2020).
39. A. E. Theyers et al., "Multisite Comparison of MRI Defacing Software Across Multiple Cohorts," Front. Psychiatry **12**, Frontiers (2021) [doi:10.3389/fpsyt.2021.617997].
40. A. F. Agarap, "Deep Learning using Rectified Linear Units (ReLU)," arXiv:1803.08375, arXiv (2019) [doi:10.48550/arXiv.1803.08375].
41. D. P. Kingma and J. Ba, "Adam: A Method for Stochastic Optimization," arXiv:1412.6980, arXiv (2017) [doi:10.48550/arXiv.1412.6980].
42. W. R. Crum, O. Camara, and D. L. G. Hill, "Generalized overlap measures for evaluation and validation in medical image analysis," IEEE Trans. Med. Imaging **25**(11), 1451–1461 (2006) [doi:10.1109/TMI.2006.880587].
43. C. H. Sudre et al., "Generalised Dice overlap as a deep learning loss function for highly unbalanced segmentations," pp. 240–248 (2017) [doi:10.1007/978-3-319-67558-9_28].
44. neuroneural, "neuroneural/brainchop," 2024, <https://github.com/neuroneural/brainchop> (accessed 11 December 2024).
45. B. BBillot, "BBillot/SynthSeg," 2024, <https://github.com/BBillot/SynthSeg> (accessed 10 December 2024).
46. A. Klein et al., "Mindboggling morphometry of human brains," PLOS Comput. Biol. **13**(2), D. Schneidman, Ed., e1005350 (2017) [doi:10.1371/journal.pcbi.1005350].
47. Kevin Wang et al., "Catalyst.Neuro: A 3D Brain Segmentation Pipeline for MRI," in Medium: PyTorch (2021).
48. J. P. Dmochowski et al., "Optimized multi-electrode stimulation increases focality and intensity at target," J. Neural Eng. **8**(4), 046011 (2011) [doi:10.1088/1741-2560/8/4/046011].
49. J. P. Dmochowski et al., "Targeted transcranial direct current stimulation for rehabilitation after stroke," NeuroImage **75**, 12–19 (2013) [doi:10.1016/j.neuroimage.2013.02.049].
50. Y. Huang et al., "Optimized tDCS for Targeting Multiple Brain Regions: An Integrated Implementation," in 2018 40th Annual International Conference of the IEEE Engineering in Medicine and Biology Society (EMBC), pp. 3545–3548 (2018) [doi:10.1109/EMBC.2018.8513034].
51. Y. Zhang, M. Brady, and S. Smith, "Segmentation of brain MR images through a hidden Markov random field model and the expectation-maximization algorithm," IEEE Trans. Med. Imaging **20**(1), 45–57 (2001) [doi:10.1109/42.906424].
52. H. Wang and P. Yushkevich, "Multi-atlas segmentation with joint label fusion and corrective learning—an open source implementation," Front. Neuroinformatics **7**, Frontiers (2013) [doi:10.3389/fninf.2013.00027].
53. B. Menze et al., "The Multimodal Brain Tumor Image Segmentation Benchmark (BRATS)," IEEE Trans. Med. Imaging **34**(10), 1993 (2014) [doi:10.1109/TMI.2014.2377694].
54. E. Giacomello, D. Loiacono, and L. Mainardi, "Brain MRI Tumor Segmentation with Adversarial Networks," in 2020 International Joint Conference on Neural Networks (IJCNN), pp. 1–8 (2020) [doi:10.1109/IJCNN48605.2020.9207220].
55. M. C. de Verdier et al., "The 2024 Brain Tumor Segmentation (BraTS) Challenge: Glioma Segmentation on Post-treatment MRI," arXiv:2405.18368, arXiv (2024) [doi:10.48550/arXiv.2405.18368].

56. Sage Bionetworks, "BraTS 2024," 2024, <https://www.synapse.org/Synapse:syn53708249> (accessed 27 December 2024).
57. M. A. Mazurowski et al., "Deep learning in radiology: An overview of the concepts and a survey of the state of the art with focus on MRI," J. Magn. Reson. Imaging **49**(4), 939–954 (2019) [doi:10.1002/jmri.26534].
58. A. S. Lundervold and A. Lundervold, "An overview of deep learning in medical imaging focusing on MRI," Z. Für Med. Phys. **29**(2), 102–127 (2019) [doi:10.1016/j.zemedi.2018.11.002].
59. R. R. Cruces et al., "Micapipe: A pipeline for multimodal neuroimaging and connectome analysis," NeuroImage **263**, 119612 (2022) [doi:10.1016/j.neuroimage.2022.119612].
60. D. C. Van Essen et al., "The WU-Minn Human Connectome Project: An overview," NeuroImage **80**, 62–79 (2013) [doi:10.1016/j.neuroimage.2013.05.041].
61. R. C. Petersen et al., "Alzheimer's Disease Neuroimaging Initiative (ADNI)," Neurology **74**(3), 201–209 (2010) [doi:10.1212/WNL.0b013e3181cb3e25].
62. B. Fischl, "FreeSurfer," NeuroImage **62**(2), 774–781, Elsevier Science, Netherlands (2012) [doi:10.1016/j.neuroimage.2012.01.021].
63. A. Fedorov et al., "Almost instant brain atlas segmentation for large-scale studies," arXiv:1711.00457, arXiv (2017) [doi:10.48550/arXiv.1711.00457].
64. A. Fedorov et al., "End-to-end learning of brain tissue segmentation from imperfect labeling," arXiv:1612.00940, arXiv (2017) [doi:10.48550/arXiv.1612.00940].
65. A. Klein and Jason Tourville, "Mindboggle: Automated Brain Labeling and Morphometry," 2012, <https://mindboggle.info/data> (accessed 1 January 2025).
66. B. Billot et al., "Robust machine learning segmentation for large-scale analysis of heterogeneous clinical brain MRI datasets," Proc. Natl. Acad. Sci. **120**(9), e2216399120 (2023) [doi:10.1073/pnas.2216399120].
67. F. Isensee et al., "nnU-Net: a self-configuring method for deep learning-based biomedical image segmentation," Nat. Methods **18**(2), 203–211 (2021) [doi:10.1038/s41592-020-01008-z].
68. D. A. Weiss et al., "Automated multiclass tissue segmentation of clinical brain MRIs with lesions," NeuroImage Clin. **31**, 102769 (2021) [doi:10.1016/j.nicl.2021.102769].
69. R. E. Gabr et al., "Brain and lesion segmentation in multiple sclerosis using fully convolutional neural networks: A large-scale study," Mult. Scler. J. **26**(10), 1217–1226, SAGE Publications Ltd STM (2020) [doi:10.1177/1352458519856843].
70. B. Fischl et al., "Whole Brain Segmentation: Automated Labeling of Neuroanatomical Structures in the Human Brain," Neuron **33**(3), 341–355, Elsevier (2002) [doi:10.1016/S0896-6273(02)00569-X].
71. B. A. Landman et al., "Multi-parametric neuroimaging reproducibility: A 3-T resource study," NeuroImage **54**(4), 2854–2866 (2011) [doi:10.1016/j.neuroimage.2010.11.047].

# Supplement

**Training of a consensus layer**

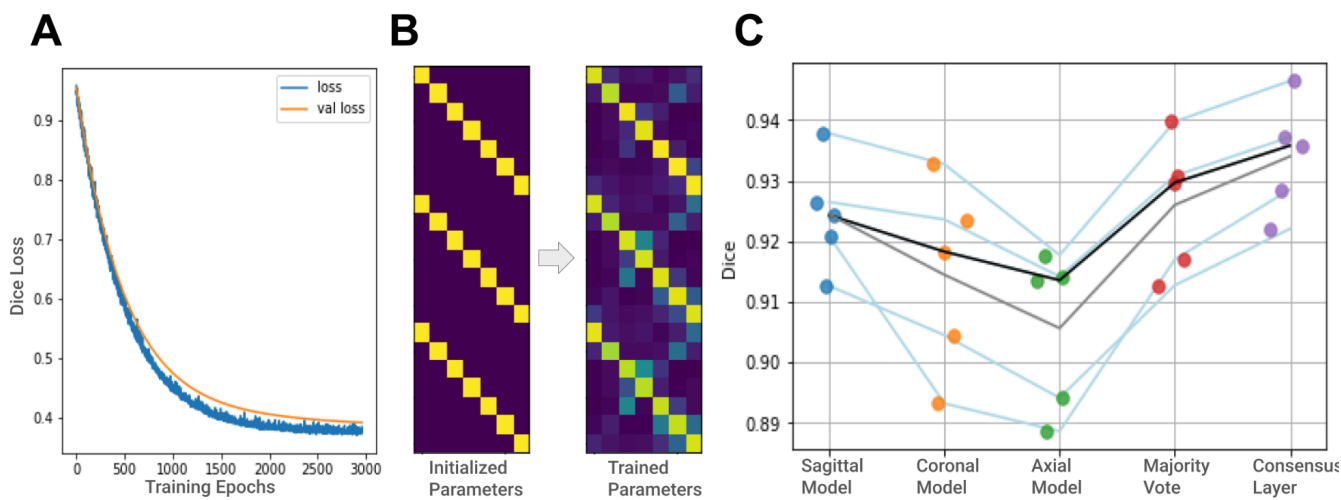

**Figure S1: Development and training of a consensus layer.** For merging the outputs of the models operating at each orthogonal direction (sagittal, coronal and axial) we use a convolutional layer that takes as input each model's output probabilities and returns a single probability per voxel for the whole MRI. A: This layer was trained for 3000 epochs with probabilities from the model's training set, with a random validation set of five MRIs for tracking performance. B: The layer was initialized by setting its weights to a diagonal matrix without bias, such that the starting point is equivalent to a majority vote, ensuring no worse performance than a simple voting method. After training the model learns to take into account alternative predictions in the inputs (off-diagonal weights). C: In the validation set, ensemble models (i.e majority vote and consensus layer) outperform each individual model. The consensus layer outperforms the majority vote (black line=median dice, gray line=average dice).

**Table S1: Median and Mean Values for Model Selection.** This table summarizes the results from Fig. 2, comparing all models during the selection process. Dice scores were calculated across all seven segmentation classes, providing both median and mean values for each model. The models are categorized as either Uniaxial (Axial, Coronal, Sagittal) or Multiaxial (Vote or Layer). Key attributes include the use of spatial information (Spa), data augmentation (Aug), and model complexity, represented by the number of base filters and depth, with larger values indicating more parameters. *In Model selection steps B and C, the Vote model (BaseFilters16_Depth6, SpaTrue_AugTrue) is the same as the Vote model from step D, but it performs worse than the **Layer Consensus model, which was ultimately chosen as the final selection.

| Model | Median Dice ± IQR | Mean Dice ± Std |
|---|---|---|
| Axial - SpaFalse_AugFalse | 0.872 ± 0.021 | 0.872 ± 0.017 |
| Axial - SpaFalse_AugTrue | 0.872 ± 0.027 | 0.877 ± 0.017 |
| Coronal - SpaFalse_AugFalse | 0.866 ± 0.026 | 0.873 ± 0.019 |
| Coronal - SpaFalse_AugTrue | 0.880 ± 0.033 | 0.879 ± 0.022 |
| Sagittal - SpaFalse_AugFalse | 0.866 ± 0.024 | 0.872 ± 0.019 |
| Sagittal - SpaFalse_AugTrue | 0.883 ± 0.023 | 0.884 ± 0.018 |
| Vote - SpaFalse_AugFalse | 0.884 ± 0.024 | 0.888 ± 0.017 |
| Vote - SpaFalse_AugTrue | 0.890 ± 0.025 | 0.894 ± 0.016 |
| Axial - SpaFalse_AugTrue | 0.872 ± 0.027 | 0.877 ± 0.017 |
| Axial - SpaTrue_AugTrue | 0.874 ± 0.028 | 0.880 ± 0.017 |
| Coronal - SpaFalse_AugTrue | 0.880 ± 0.033 | 0.879 ± 0.022 |
| Coronal - SpaTrue_AugTrue | 0.884 ± 0.035 | 0.883 ± 0.022 |
| Sagittal - SpaFalse_AugTrue | 0.883 ± 0.023 | 0.884 ± 0.018 |
| Sagittal - SpaTrue_AugTrue | 0.879 ± 0.026 | 0.885 ± 0.017 |
| Vote - SpaFalse_AugTrue | 0.890 ± 0.025 | 0.894 ± 0.016 |
| Vote - BaseFilters8_Depth6 | 0.887 ± 0.022, | 0.889±0.016 |
| Vote - BaseFilters4_Depth6 | 0.871 ± 0.020 | 0.876±0.016 |
| Vote - BaseFilters16_Depth5 | 0.890 ± 0.020 | 0.891±0.018 |
| Vote - BaseFilters16_Depth4 | 0.885 ± 0.028 | 0.884±0.021 |
| Vote - BaseFilters32_Depth7 | 0.899 ± 0.029 | 0.900±0.017 |
| Vote - BaseFilters16_Depth6 SpaTrue_AugTrue* | 0.893 ± 0.027 | 0.896 ± 0.016 |
| **Layer Consensus**** | **0.895 ± 0.024** | **0.899 ± 0.016** |

**Table S2: Statistical Results of Multiaxial Full Head Segmentation.** This table presents the results from Fig. 3A, which compares Multiaxial segmentation with other full head segmentation methods on Table 1 dataset. All seven classes were considered for the Dice scores. It includes the median and mean values, along with statistical significance.

| Dice Score | | | Statistical Analysis | | | |
|---|---|---|---|---|---|---|
| Model | Median ± IQR | Mean ± STD | Comparisons | Test | Statistic | p |
| Multiaxial | 0.8724 ± 0.0505 | 0.8662 ± 0.0363 | Group | Friedman Test | F=130 | 4.90e-28 |
| Multipriors SPM8 | 0.8631 ± 0.0398 | 0.8549 ± 0.0364 | Multiaxial vs Multipriors SPM8 | Wilcoxon Test | W=279 | 1.11e-04 |
| Multipriors SPM12 | 0.84 ± 0.0843 | 0.8133 ± 0.0713 | Multipriors SPM8 vs Multipriors SPM12 | Wilcoxon Test | W=24 | 9.26e-10 |
| SPM12 | 0.7891 ± 0.1028 | 0.7486 ± 0.1102 | Multipriors SPM12 vs SPM12 | Wilcoxon Test | W=47 | 3.26e-09 |
| SynthSeg Retrained | 0.7611 ± 0.0302 | 0.7599 ± 0.0273 | SPM12 vs SynthSeg Retrained | Wilcoxon Test | W=593 | 3.82e-09 |

**Table S3: Statistical Results of Multiaxial Gray and White Matter Brain Segmentation.**
This table presents the results from Fig. 3B, which compares Multiaxial segmentation with other GWM segmentation methods on Table 1 dataset. Only GWM Dice scores were considered. It includes the median and mean values, along with statistical significance.

| Dice Score | | | Statistical Analysis | | | |
|---|---|---|---|---|---|---|
| Model | Median ± IQR | Mean ± STD | Comparisons | Test | Statistic | p |
| Multiaxial | 0.8862 ± 0.0796 | 0.8689 ± 0.06 | Group | Friedman Test | F=102 | 6.58e-23 |
| SynthSeg | 0.805 ± 0.0654 | 0.7952 ± 0.0486 | Multiaxial vs SynthSeg | Wilcoxon Test | W=4 | 3.00e-10 |
| Brainchop | 0.7599 ± 0.073 | 0.7482 ± 0.0549 | SynthSeg vs Brainchop | Wilcoxon Test | W=0 | 2.37e-10 |

**Table S4: Statistical Results of Multiaxial Brain Segmentation on Mindboggle Dataset**
This table presents the results from Fig. 4, which compares Multiaxial segmentation with other GWM segmentation methods compared to the Mindboggle dataset. Only GWM Dice scores were considered. It includes the median and mean values, along with statistical significance.

| Dice Scoring | | | | Statistical Analysis | | | |
|---|---|---|---|---|---|---|---|
| Model | Dataset | Median ± IQR | Mean ± STD | Comparisons | Test | Statistic | p |
| Multiaxial | NKI | 0.9076 ± 0.018 | 0.907 ± 0.0159 | Group | Friedman Test | F=48 | 4.68e-11 |
| Brainchop | NKI | 0.8963 ± 0.009 | 0.8944 ± 0.0098 | Multiaxial vs Brainchop | Wilcoxon Test | W=129 | 1.94e-05 |
| SynthSeg | NKI | 0.9165 ± 0.0095 | 0.9162 ± 0.0066 | Brainchop vs SynthSeg | Wilcoxon Test | W=0 | 4.55e-13 |
| Multiaxial | OASIS/MMRR | 0.8191 ± 0.0324 | 0.8176 ± 0.0252 | Group | Friedman Test | F=80 | 4.15e-18 |
| Brainchop | OASIS/MMRR | 0.882 ± 0.0353 | 0.8926 ± 0.0186 | Multiaxial vs Brainchop | Wilcoxon Test | W=1 | 1.82e-12 |
| SynthSeg | OASIS/MMRR | 0.9212 ± 0.0166 | 0.9204 ± 0.0098 | Brainchop vs SynthSeg | Wilcoxon Test | W=0 | 9.10e-13 |

**Table S5: Statistical Results of Multiaxial Using Original, Defaced, and Refaced MRIs.**
This table presents the results from Fig. 5, which compares Multiaxial segmentation using different defacing methods on Table 1 dataset. Only GWM Dice scores were considered. It includes the median and mean values, along with statistical significance.

| | Dice Score | | Statistical Analysis | | | |
|---|---|---|---|---|---|---|
| MRI Type | Median ± IQR | Mean ± STD | Comparisons | Test | Statistic | p |
| Original | 0.8978 ± 0.0412 | 0.891 ± 0.0386 | Group | Friedman Test | F=50.5 | 1.07e-11 |
| Defaced | 0.8866 ± 0.052 | 0.8831 ± 0.0415 | Original vs Defaced | Wilcoxon Test | W=61 | 6.87e-09 |
| Refaced | 0.8926 ± 0.0515 | 0.8841 ± 0.0407 | Original vs Refaced | Wilcoxon Test | W=512 | 7.16e-02 |

**Table S6: Statistical Results of Current-Flow Model Comparison**
This table presents the statistical results corresponding to Fig. 6B, comparing current-flow models based on RMS differences in target electric field magnitude relative to the ground truth. The four methods compared were: Multipriors segmentation with SPM12 alignment, Multiaxial segmentation with SPM12 alignment, Multiaxial segmentation with manual alignment, and Manual Ground Truth Labels with manual alignment. Pairwise comparisons assess the impact of segmentation quality and alignment accuracy on model performance.

| | Root Mean Square Error | | Statistical Analysis | | | |
|---|---|---|---|---|---|---|
| Method | Median ± IQR | Mean ± STD | Comparison | Test | Statistic | p |
| Multiaxial Manual | 0.0555 ± 0.0183 | 0.0553 ± 0.0251 | Group | Friedman Test | F=12.8 | 0.0017 |
| Multiaxial SPM12 | 0.0678 ± 0.0391 | 0.0763 ± 0.0345 | Multiaxial Manual vs Multiaxial SPM12 | Wilcoxon Test | W=8 | 0.048 |
| Multipriors SPM12 | 0.0957 ± 0.0159 | 0.0974 ± 0.0275 | Multiaxial Manual vs Multipriors SPM12 | Wilcoxon Test | W=0 | 0.002 |

**Instructions on how to run the demo code for new Kaggle users:**

https://www.kaggle.com/code/andrewbirnbaum/multiaxial-demo.

To run this demo you must be signed into Kaggel with a verified account. Select Copy & Edit on the top left and select "Edit in Kaggle". In Session Options (bottom right) select "Internet On", and execute the first cell which installs the latest version of Tensorflow. Then select "Restart and run all cells" (three dots on top right). Select a subject number when prompted.

Data and model has been released for open use, using license Creative Commons Attribution-NonCommercial-ShareAlike 4.0 International.

The ROAST current-flow modeling pipeline begins with MRI preprocessing steps, including isotropic resampling, RAS orientation, alignment, and segmentation. Our Multiaxial method replaces the alignment and segmentation steps, as it does not depend on prior alignment. This allows for straightforward integration into the existing ROAST pipeline. Because our segmentation model is implemented in TensorFlow, Python must be accessed from within MATLAB, where ROAST operates. To facilitate this, we implemented a Bash script that sets up a Python virtual environment, enabling seamless communication between MATLAB and Python. Importantly, this process is automated, and users do not need to modify their typical ROAST workflow. Matlab calls a single python script to execute the segmentation taking a nii file as input and output. For cases requiring electrode adjustment, an optional manual alignment GUI is available (see Fig. S2). Beyond this point, the ROAST pipeline proceeds as usual.

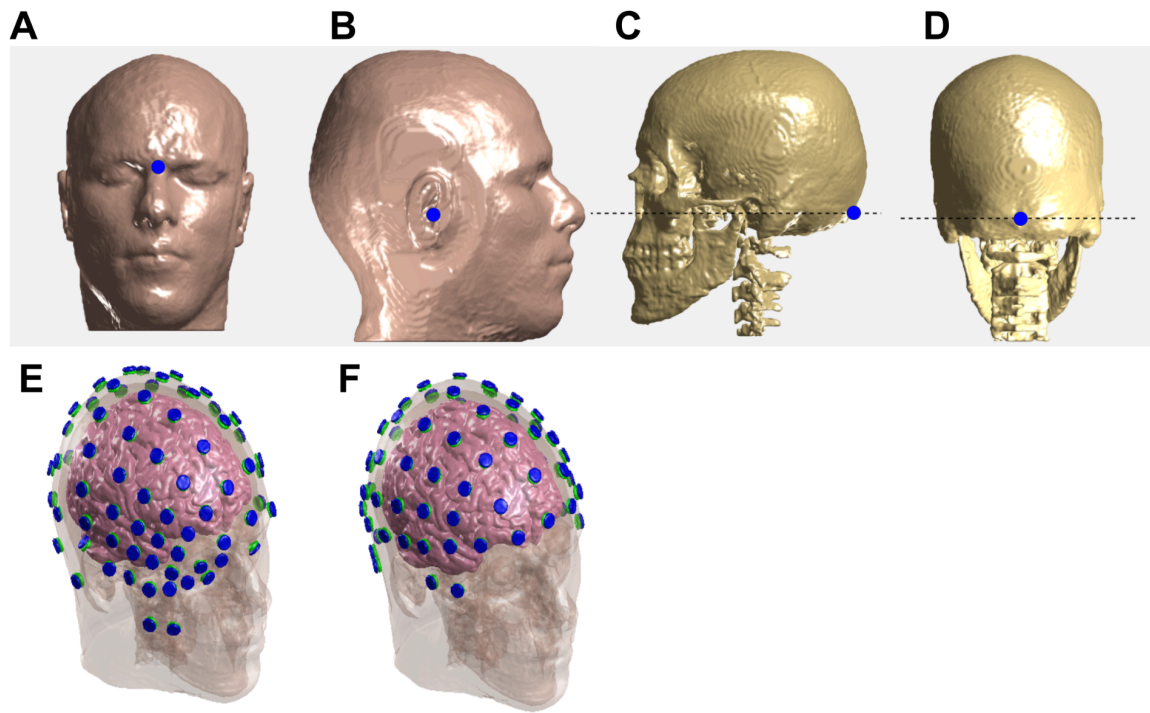

**Figure S2: Electrode Placement and Landmark Selection in ROAST A:** Shows the nasion. **B:** Shows the right ear (the left ear is placed the same way). **C-D:** Shows how to properly select the inion, which cannot be seen on the skin, so the skull is used for placement. **E-F**: (Blue represents transcranial electrodes, green represents conductive electrolyte gel) **E:** Shows misaligned electrodes due to improper landmark placement. **F**: Shows proper electrode placement with manually selected landmark positions matching the International 10–20 system.

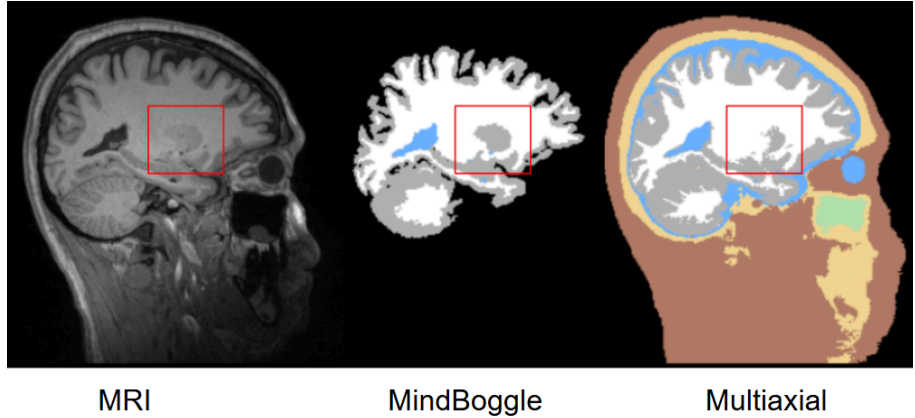

**Figure S3: Mixed gray and white matter in subcortical parcellation.** The basal ganglia in MRI (dark area in the red rectangle) appears as both gray and white but is labeled in the MindBoggle data as a single parcel (here it is shown as gray). In our manual ground truth it is labeled as both gray and white matter and therefore segmented as such by the Multiaxial model.